\documentclass[aps,prl,twocolumn,amsmath,amssymb,floatfix,
superscriptaddress]{revtex4}
\usepackage{tabularx}
\usepackage{bm}
\usepackage{euscript}
\usepackage{epsfig,psfrag,subfigure}
\usepackage{graphicx}
\usepackage{color}
\usepackage{amsfonts}
\usepackage{exscale}
\usepackage{amsbsy}

\usepackage{pstricks}
\newlength{\myL}

\def\avg#1{\left\langle#1\right\rangle}
\def\bra#1{\left\langle#1\right|}
\def\ket#1{\left|#1\right\rangle}
\def\braket#1#2{\left\langle #1\right|\left.#2\right\rangle}

\def\be{\begin{equation}}       \def\ee{\end{equation}}
\def\bea{\begin{eqnarray}}      \def\eea{\end{eqnarray}}
\def\ba{\begin{array} }
\def\ea{\end{array} }
\def\bnum{\begin{enumerate} }
\def\enum{\end{enumerate}}

\def\nn{\nonumber}

\def\=>{\Rightarrow}
\def\>{\rightarrow}
\def\A{\uparrow}
\def\V{\downarrow}

\def\eye2{Fathbb{I}}

\def\Fig#1{Fig.~\ref{#1}}

\begin{document}
\title{Symmetry fractional quantization in two dimensions}
\author{Hong Yao}
\affiliation{Department of Physics, University of California, Berkeley, CA, 94720}
\affiliation{Materials Sciences Division,
Lawrence Berkeley National Laboratory, Berkeley, CA 94720, USA}
\author{Liang Fu}
\affiliation{Department of Physics, Harvard University, Cambridge, MA, 02138}
\author{Xiao-Liang Qi}
\affiliation{Department of Physics, Stanford University, Stanford, CA, 94305}
\date{\today}
\begin{abstract}
We introduce a solvable spin-rotational and time-reversal invariant spin-1 model in two dimensions. 
Depending on parameters, the ground state is an equal-weight superposition of all
valence loops called ``resonating valence loop'' (RVL) or an equal-weight superposition of all
valence bonds known as ``resonating valence bond'' (RVB). In RVL, ends of open loops are deconfined spinons of spin-1/2 that cannot be obtained by simple combinations of spin-1 -- a phenomenon of fractionalization; while in RVB, all quasiparticles carry an integer spin, hence 
no fractionalization.  
 RVL and RVB are spin liquids with {\it identical} topological order but {\it different} spin-rotational and time-reversal symmetry
quantum numbers of quasiparticles.
We propose that quantized symmetry
quantum number gives a systematic way to (partially) classify phases with identical topological order in dimensions greater than one.
\end{abstract}
\maketitle

Quantum number fractionalization\cite{kivelson:sm02} is among the most striking quantum phenomena in condensed matter physics: quasiparticles can have fractional quantum number that cannot be obtained by simple combination of fundamental constituents of the system. 
In most known examples of dimensions greater than one, fractional quantum numbers are {\it quantized} to certain discrete values and the origin of fractional quantization comes from topological order\cite{topo-order} of the underlying system. 
For instance, topological order is solely responsible for the quantization of fractional charge of quasiparticles in fractional quantum Hall (FQH) states, whereas the $U(1)$  charge symmetry allows, a priori, continuously tunable fractional charge\cite{irrational-charge}. We call such phenomenon ``{\it topological fractional quantization}''. 

There are also cases where the symmetry property by itself dictates quantization of fractional quantum numbers. This occurs when the symmetry group has only {\it discrete} irreducible (projective) representations.  
We call such phenomenon ``{\it symmetry fractional quantization}''. For both cases of topological and symmetry fractional quantization, it is possible to realize two distinct states exhibiting {\it identical} topological order but supporting quasiparticles with {\it different} quantized fractional quantum numbers. 

To illustrate symmetry fractional quantization, we introduce an exactly solvable $SO(3)$
spin-rotational and time-reversal invariant spin-1 model on the (decorated) honeycomb lattices. Depending on parameters in the model,
the ground state is an equal-weight superposition of valence loops called ``resonating valence loop'' (RVL) or an equal-weight superposition of valence bonds known as ``resonating valence bond'' (RVB)\cite{anderson:mrb73,kivelson}. The RVL and RVB are $Z_2$ spin liquids with identical topological order. But they are distinct phases of matter because their quasiparticles carry different quantum number quantized by symmetry.
Specifically, in RVL, each loop is like an 
AKLT spin-1 chain \cite{affleck:prl87} and the ground state is a loop soup pictorially. Ends of open loops
are deconfined spinons 
with spin-1/2. Spin-1/2 objects form a {\it projective representation} of $SO(3)$ group which cannot be obtained from any combination (tensor product) of integer spin representations\cite{projective}. Moreover, spin-1/2 transforms differently from spin-1 under time reversal operation $\cal T$: ${\cal T}^2=1$ for spin-1, whereas ${\cal T}^2=-1$ for spin-1/2; the latter leads to Kramer's degeneracy. Thus the RVL exhibits fractionalization of quantum numbers of 
 spin-rotational and time-reversal symmetry. To the best of our knowledge, 
this is the first example of quantum number fractionalization of time-reversal symmetry realized in a 
solvable model\cite{sachdev}. In contrast, in RVB, all
quasiparticles carry integer spins, hence no fractionalization.

One motivation to study symmetry 
fractional quantization is to classify topological phases of matter with symmetry. 
For free fermion systems, the topological classification with discrete symmetries such as time-reversal symmetry and particle-hole symmetry has been studied systematically\cite{ludwig} since the recent discovery of topological insulators and superconductors\cite{kane,qi,moorereview}. More recently, the classification has been generalized to insulators with certain spatial symmetries\cite{moore, fu, turner}. For  interacting systems, the classification has been intensively studied in  1D\cite{wen1d,turner1d,kitaev1d}; however, classification in higher dimensions is far from complete (see, nonetheless, Refs.\cite{wenz2,kou,levin,yao}).  
We hereby propose that symmetry quantum numbers of quasiparticles can (partially) distinguish topologically different phases with the same symmetry. The reason is that symmetry quantum numbers are {\it discrete} quantities and cannot change smoothly as long as the underlining symmetries are respected. Therefore two gapped phases in which quasiparticles transform under different irreducible representations of a symmetry group cannot be adiabatically connected under symmetry-preserving deformations: they are necessarily different phases. 
Topological order, combined with (fractional) symmetry quantum numbers, give a more refined classification of topological phases with symmetry.
It is worth to mention that Ref.\cite{wenz2,kou} has classified hundreds of distinct $Z_2$ spin liquids of spin-1/2 systems protected by space group symmetry, based on the projective symmetry group (PSG) approach to study the variational ground state wave function obtained from slave-particle construction. The connection 
between the PSG property of {\it ground states} and the physical quantum number of {\it quasiparticles} is not perfectly  understood. Moreover, the distinction between spin liquids protected by spatial symmetry is not robust against disorder, whereas our study dealing with time-reversal symmetry is.

\begin{figure}[b]
\subfigure[]{
\includegraphics[scale=0.17]{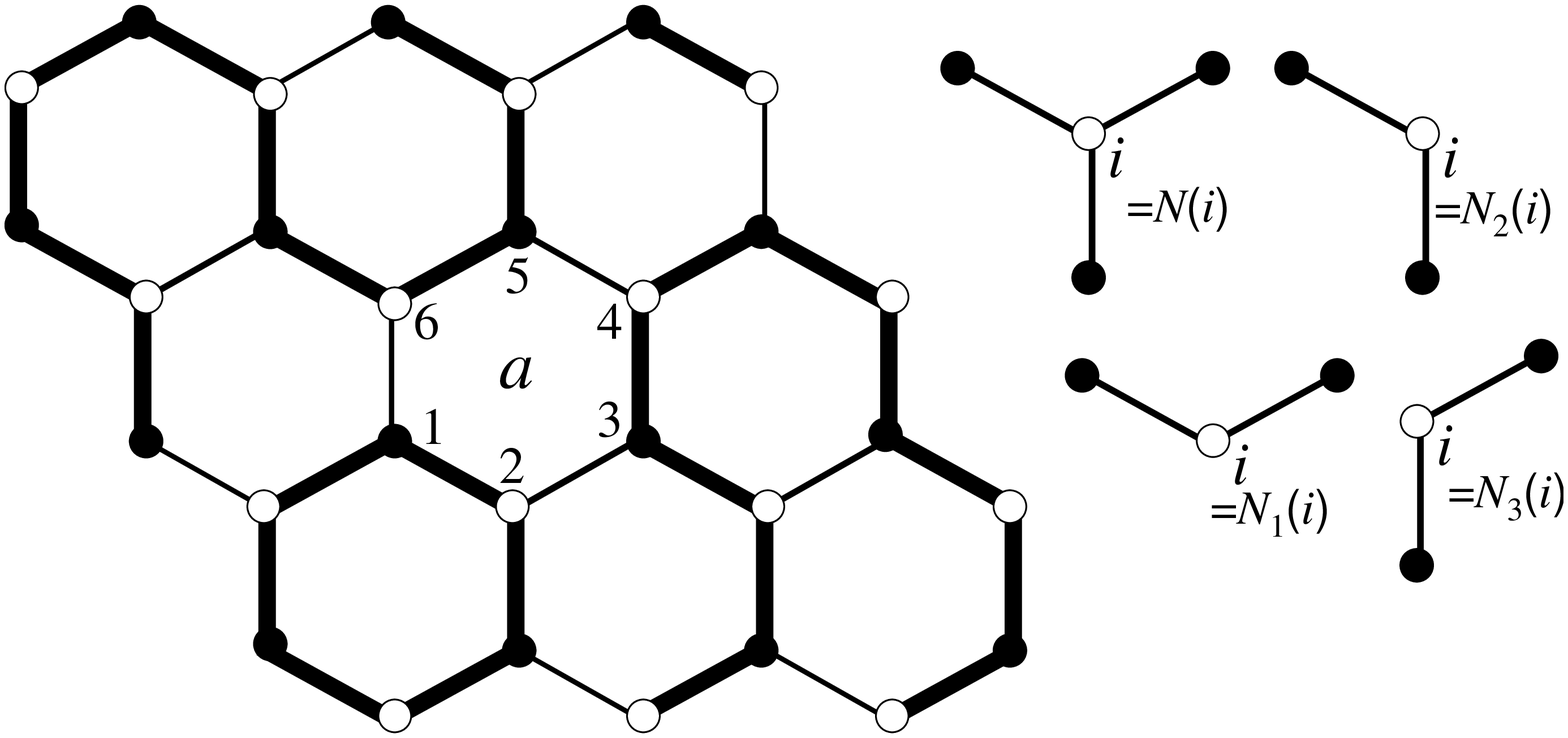}}~~~
\subfigure[]{
\includegraphics[scale=0.18]{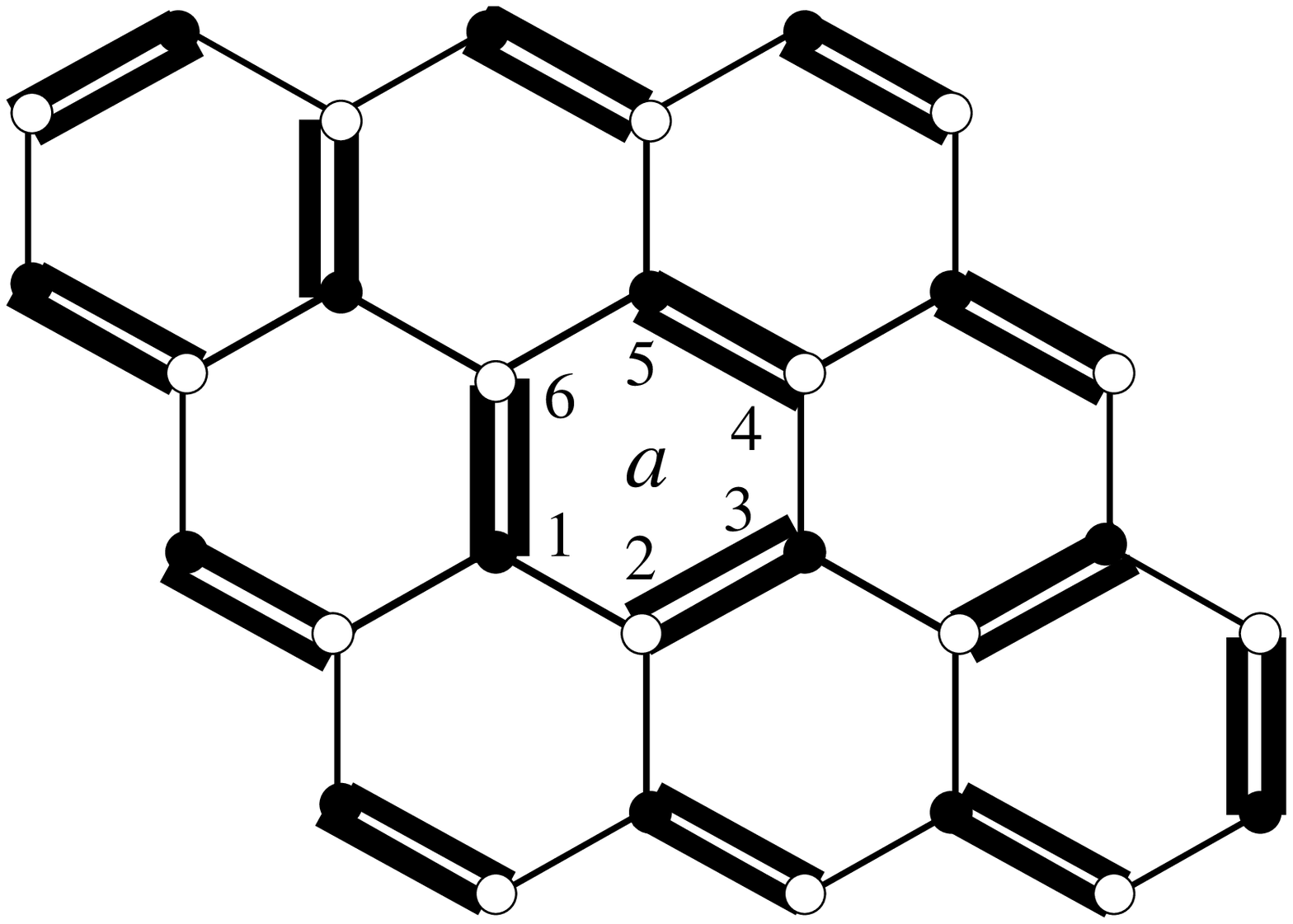}}
\caption{(a) The schematic representation of the honeycomb lattice and a typical loop-covering configuration. The black (white) sites are in A (B) sublattices. Thick bond represent a spin singlet created by $B^\dag_{ij}$ (see text)
and they forms a loop-covering on the honeycomb lattice. A hexagon plaquette is flippable if it has three thick and three thin bonds. For instance, the loop configuration is flippable on the hexagon plaquette marked by $a$. (b) A typecal spin-1 dimer-covering configuration in which each dimer consists of two singlet bonds. }
\label{fig:1}
\end{figure}

{\bf Resonating valence loops (RVL):} For simplicity, we shall consider a spin-1 system on the honeycomb lattice first. On each site $i$, $S^\alpha_i =\frac12 b^\dag_{i\mu} \sigma^\alpha_{\mu\nu} b_{i\nu}$, where $\sigma^\alpha$ are Pauli matrices and $b_{i\mu}$ are Schwinger bosons with the constraint  $b^\dag_{i\A}b_{i\A}+b^\dag_{i\V}b_{i\V}=2$. In the spirit of AKLT, each of two bosons on each site can form a singlet bond with another boson on its nearest neighbor site  - each  spin-1 can participate two singlet bonds. A spin singlet bond on the link $\avg{ij}$ is created by $B^\dag_{ij} =\epsilon^{\mu\nu}b^\dag_{i\mu}b^\dag_{j\nu}$ ($\epsilon^{\mu\nu}$ is the Levi-Civita symbol). For reasons which will be clear later, we are interested in the following states:
\bea
\ket{\Psi}_\textrm{RVL}=\sum_c\ket{c};~~ \ket{c}=(-1)^{n_c}
\prod_{\avg{ij}\in c}B^\dag_{ij} \ket 0,
\label{eq:c}
\eea

\vspace{-0.3cm}

\noindent where $c$ is a loop-covering configuration which consists of non-intercepting loops and which touches every site by one and only one loop and $n_c$ counts the number of singlet bonds of the loop configuration $c$ on those vertical links with upper site in sublattice $A$.
$\ket{\Psi}_\textrm{RVL}$ is an {\it equal weight} superposition of all loop-covering configurations, which we propose to call ``resonating valence loop'' (RVL) state. A typical loop-covering configuration is shown in \Fig{fig:1}. For the honeycomb (or any other trivalent) lattice, there is a one-to-one correspondence between loop-covering and dimer-covering configurations: for any loop-covering $c$, a corresponding dimer configuration is obtained by occupying a virtual dimer on any link not covered by the loop; consequently, the number of loop configurations is identical to the number of dimer ones that increases exponentially with the size of the lattice. Since $B^\dag_{ij}=-B^\dag_{ji}$, in Eq. (\ref{eq:c}), we take the convention  $i(j)\in A(B)$ sublattices for all links $\avg{ij}$ such that $\braket{c}{\tilde c}<0$ where $\tilde c$ is obtained from $\ket{c}$ by flipping singlet bonds which are flippable on some hexagon plaquette. The negative overlap\cite{cano:prl10} between $\ket{c}$ and $\ket{\tilde c}$ is important for reasons which will become clear later.

Before dicussing properties of the RVL state, we shall first introduce a microscopic model for which the RVL state in Eq. (\ref{eq:c}) is the ground state.
Since the RVL state consists of loop configurations, it is desired to associate an energy cost for non-loop states. To achive this, we consider
the following Klein term\cite{klein:82}
\bea
H_K&=&J\sum_{i}\bigg[P_{3}[{\cal N}(i)]+ P_4[{\cal N}(i)]+\sum_{\alpha=1}^3 P_3[{\cal N}_\alpha(i)]\bigg],~~
\label{eq:H0}
\eea

\vspace{-0.3cm}

\noindent where $J>0$, ${\cal N}(i)$ labels a cluster of four sites consisting of $i$ and three neighbors of $i$, ${\cal N}_\alpha(i)$ a cluster of three sites consisting of $i$ and two neighbors of $i$, as shown in Fig. \ref{fig:1}(a). Here $P_s[{\cal N}(i)]$ is a projection operator onto the sector of total spin-$s$ of the cluster ${\cal N}(i)$, namely
\bea
P_s[{\cal N}(i)]=\prod_{r=0,r\neq s}^{4}\frac{\vec S_{{\cal N}(i)}^2-r(r+1)}{s(s+1)-r(r+1)}
\eea

\vspace{-0.2cm}

\noindent where $\vec S_{{\cal N}(i)}=\sum_{j\in{\cal N}(i)}\vec S_j $. $P_s[{\cal N}_\alpha(i)]$ and other projection operators which will appear later are defined similarly.
Since $H_K$ is a sum of projection operators, it is positive semi-definite. It is clear that $H_K\ket c=0$;  $\ket c$ is a ground state of $H_K$. As usual, one question about the Klein term is whether it is perfect, namely whether there are non-loop configurations whose energy is zero. Although lacking a rigorous proof\cite{chayes:cmp89}, we believe that the set of $\ket c$ forms the complete ground state manifold of $H_K$ since there are no obvious non-loop states with zero energy.

The dual description in terms of dimers tells us how loop configurations resonate:
if $c$ has three separated singlet bonds on a hexagon plaquette, $c$ is said to be flippable on that plaquette, as shown in \Fig{fig:1}(a). It is then clear that the maximum possible value of the total spin $\vec S_a=\sum_{i\in a}\vec S_i$ of a hexagon $a$ is 3 when $a$ is flippable. For all other configurations non-flippable on $a$, the maximum possible spin on $a$ is at most 2. This fact makes flippable configuration unique and enables us to construct the following Hamiltonian for which we shall prove that $\ket{\Psi}_\textrm{RVL}$ is its ground state.

\vspace{-0.6cm}

\bea
H_\textrm{RVL}&=&H_K+H_Q, \nn\\
&=&H_K+J'\sum_{a}P_3(a)\left[P_3(E_{a}) +P_3(O_{a})\right],
\label{eq:H1}
\eea

\vspace{-0.5cm}

\noindent where $O_a (E_a)$ represents the odd (even) sites,  namely the three $A(B)$ sublattice sites,  of plaquette $a$. Note that $P_3(a)$, $P_3(E_a)$ and $P_3(O_a)$ commute with each other. 

It is clear that $H_\textrm{RVL}$ is positive semi-definite since it is also a sum of projection operators. Consequently, we can conclude that $\ket{\Psi}_\textrm{RVL}$ is a ground state of $H_\textrm{RVL}$ if we prove $H_\textrm{RVL}\ket{\Psi}_\textrm{RVL}=0$ or equivalently $H_Q\ket{\Psi}_\textrm{RVL}=0$ by knowing that $H_K\ket{\Psi}_\textrm{RVL}=0$. To prove this, we choose a plaquette $a$ and rewrite $
\ket{\Psi}_\textrm{RVL}=\ket{\psi_a} +\sum_{c_a} \ket{c_a}$,
where $c_a$ label loop configurations which are flippable on $a$, and $\ket{\psi_a}$ contains all other configurations non-flippable on $a$. As discussed earlier, $S_a\leq 2$ for a configuration non-flippable on $a$; this implies that $P_3(a)[P_3(E_a)+P_3(O_a)]\ket{\psi_a}=0$. For each $\ket{c_a}$,
$P_3(a)\ket{c_a}\neq0$; however, there is another configuration $\ket{\tilde c_a}$ which is related to $\ket{c_a}$ by flipping the singlet bonds on $a$.
Now we shall show that $P_3(E_a)\left(\ket{c_a}+\ket{\tilde c_a}\right)=0$ and the same for $P_3(O_a)$.
The projection $P_3(E_a)$ can be written in spin-$3$ coherent states
\bea
P_3(E_a)=\frac 7{4\pi}\int{d^2{\bf \hat{n}}}\ket{S_{E_a}=3,\hat{\bf n}}\bra{S_{E_a}=3,\hat{\bf n}},
\eea
where $\ket{S_{E_a}=3,\hat{\bf n}}$ is the state with  $S_{E_a}=3$ and the spin component ${\bf S}_{E_a}\cdot{\bf\hat{n}}=3$ along the unit vector $\hat{\bf n}$. Define $P_3^{\bf \hat{n}}(E_a)=\ket{S_{E_a}=3,\hat{\bf n}}\bra{S_{E_a}=3,\hat{\bf n}}$ as the projector to the $\hat{\bf n}$ direction. Due to spin rotation invariance of the loop states $\ket{c_a}$ and $\ket{\tilde{c}_a}$, one can prove $P_3^{\hat{\bf n}}(E_a)\left({\ket{c_a}+\ket{\tilde{c}_a}}\right)=0$ for all $\hat{\bf n}$ if it's true for a particular $\hat{\bf n}$. Thus we take $\hat{\bf n}=\hat{\bf z}$ and study the action of $P_3^{\hat{\bf z}}(E_a)$. From  $(\ket{c_a}+\ket{\tilde c_a})
=\left[\prod_{\avg{ij}\in c_a, \avg{ij}\notin a} B^\dag_{ij}\right]\left[B^\dag_{12}B^\dag_{34} B^\dag_{56}-B^\dag_{16}B^\dag_{32}B^\dag_{54}\right] \ket{0}$, it is clear that the contribution from $\left[\prod_{\avg{ij}\in c_a, \avg{ij}\notin a} B^\dag_{ij}\right]$ to $S^z(E_a)=S^z_2+S^z_4+S^z_6$ is at most 3/2 while the contribution from $\left[B^\dag_{12}B^\dag_{34} B^\dag_{56}-B^\dag_{16}B^\dag_{32}B^\dag_{54}\right]$ is at most 1/2; consequently,  $S^z(E_a)\leq 2$ for $(\ket{c_a}+\ket{\tilde c_a})$, which implies that $P^{\bf \hat z}_3(E_a)(\ket{c_a}+\ket{\tilde c_a})=0$. Consequently,
\bea
H_\textrm{RVL}\ket{\Psi}_\textrm{RVL}=(H_K+H_Q)\ket{\Psi}_\textrm{RVL}=0.
\eea
Thus, we have proved that $\ket{\Psi}_\textrm{RVL}$ is the exact ground state of $H_\textrm{RVL}$.
Since $P_3(a)[P_3(E_a)+P_3(O_a)](\ket{c_a}-\ket{\tilde c_a})\neq 0$, we believe that the ground state of $H_\textrm{RVL}$ in each topological sector is unique.
Because the honeycomb lattice is bipartite, by its analogy to the quantum dimer model\cite{rokhsar:prl88} the RVL state on the honeycomb lattice is expected to describe a critical point rather than a stable phase - small perturbations would drive it into a different state\cite{moessner:prb01}.  The exact nature of states at and around this critical point remains to be studied in future.

{\bf Decorated honeycomb lattice:} To obtain a stable spin liquid phase described by the RVL  state, we consider a spin-1 system on the decorated honeycomb lattice (see Fig. 1 in Ref. \cite{yao:prl07}), also known as the star or the 3-12 lattice. The RVL state here is formally the same as in Eq. (\ref{eq:c}).
We consider the following Hamiltonian that is a direct extention of the one studied on honeycomb lattice:
\bea
H_\textrm{RVL}=H_K+J'\sum_a P_6(a)[P_6(E_a)+P_6(O_a)],
\label{eq:Hdeco}
\eea
where $H_K$ is the Klein term which is identical to the one in Eq. (\ref{eq:H0}). Here $a$ labels a dodecagon  instead of hexagon plaquette. It is straightfoward to prove that $\ket{\Psi}_\textrm{RVL}$ on the decorated honeycomb lattice is the ground state of $H_\textrm{RVL}$ in Eq. (\ref{eq:Hdeco}).
Since the decorated honeycomb lattice is non-bipartite, by analogy to quantum dimer model\cite{moessner:prl01}, the short-range RVL state is gapped and then stable against small perturbations. We expect that there is a finite region in parameter space around the Hamiltonian Eq. (\ref{eq:Hdeco}) such that the ground state is a loop-liquid state with a finite gap to all excited states. From now on, we will focus on the decorated honeycomb lattice unless stated othervise.

{\bf Deconfined spinons:} In the ground state, all loops are closed in each $\ket{c}$. What happens when some loops are open? Since each loop is like a spin-1 AKLT chain, from the physics of 1D AKLT spin-1 chain, it is clear that spin-1/2 excitations would appear at the ends of open loops. These spin-1/2 excitations are so-called spinons. (In the dual-dimer picture, a spinon excitation occurs at the site where two different dimers overlap.)   The wave function with two static spinons can be written down explicitly\cite{spinon}: 
\bea
\ket{\Psi_{i\mu,j\nu}}=b^\dag_{i\mu} b^\dag_{j\nu}\sum_{c_{ij}}\prod_{\avg{kl}\in c_{ij}} B^\dag_{kl}\ket{0},
\eea
where $c_{ij}$ label configurations with only one open loop whose ends are $i$ and $j$. It is clear that $\ket{\Psi_{i\mu,j\nu}}$ has two spin-1/2 excitations localized at $i$ and $j$, respectively, and each of them is double degenerate.  The energy associated with the two spinon excitations is finite since the constraint imposed by the projection operators is only violated around $i$ and $j$ locally evenwhen they are far separated. Specifically, the energy cost is at most $4J$ since only $P_3[{\cal N}(i)]$ and one of $P_3[{\cal N}_\alpha(i)]$ cost finite energy around $i$ and similarly around $j$. Consequently, deconfined spinons are supported
in the RVL liquid phase on the decorated honeycomb lattice.

Each spinon excitation carries spin-1/2 quantum number of the $SO(3)$ spin-rotational symmetry and spin-1/2 cannot be obtained from simple combinaitons of fundamental constitutents that carry spin-1. Moreover, the time-reversal transformation ${\cal T}$ of spinons satisfies ${\cal T}^2=-1$.
Since the fundamental constituents are spin-1 which have ${\cal T}^2=1$, spinon excitations exhibit time reversal symmetry quantum number fractionalization. 
As an important consequence, even when the model is perturbed by spin $SO(3)$ non-invariant terms, the double degeneracy of  ``spinon'' excitations would remain due to the Kramers degeneracy, as long as time-reversal symmetry is preserved.   

{\bf Resonate valence bonds (RVB):} To compare the RVL state with other liquid states and obtain better understanding on the symmetry fractional quantization, we study a spin-1 RVB state which is a direct generalization of the spin-1/2 RVB state given by
\bea
\ket{\Psi}_\textrm{RVB}=\sum_c\ket{c};~~ \ket{c}= (-1)^{n_c}\prod_{\avg{ij}\in c} (B^\dag_{ij})^2\ket{0},
\eea

\vspace{-0.2cm}

\noindent where $c$ labels a dimer-covering and $n_c$ is chosen similarly such that $\braket{c}{\tilde c}<0$. The spin on each site forms a singlet with one of its neighbors in this spin-1 RVB state. With some straightforward calculation, it is clear that this spin-1 RVB state is a ground state of the following Hamiltonian with short-range interactions
\bea
H_\textrm{RVB}=J\sum_i \left[ \sum_{s=3}^4 P_s[ {\cal N}(i)]+P_2[{\cal N}(i)]\sum_{\alpha=1}^3 P_0[{\cal N}_\alpha(i)]  \right]\nn\\
+J'\sum_a \left[ P_6(E_a) \prod_{i\in O_a} P_2(ii')+P_6(O_a) \prod_{i\in E_a} P_2(ii') \right], ~
\eea

\vspace{-0.1cm}

\noindent where $ii'$ label the link not in the hexagon plaquette $a$. Again, by analogy with quantum dimer model\cite{moessner:prl01}, we believe that this Hamiltonian is gapped and the spin-1 RVB state is the unique ground state in each topological sector.
The RVB and RVL states are both $Z_2$ spin liquids with identical  topological order. However, there is a qualitative distinction between the two states: spin-1/2 (spinon)  excitations in the spin-1 RVB phase is linearly confined while they are deconfined in the RVL phase.

{\bf Phase transition(s) between RVL and RVB:} When the time-reversal, $SO(3)$ spin-rotation, or both symmetries are respected, spinons exhibit fractional quantization of time reversal or $SO(3)$ quantum numbers; consequently it is impossible to adiabatically connect the RVL state with the RVB without going through a phase transition even though they have identical topological order. To study the possible phase transition(s) between them, we introduce the Hamiltonian
$H(\lambda)=\lambda H_\textrm{RVB}+(1-\lambda) H_\textrm{RVL}$.
By tuning $\lambda$ from 0 to 1, we expect at least one phase transition between the RVL phase at $\lambda=0$ and the RVB phase at $\lambda=1$.
Nonetheless, when both time reversal and $SO(3)$ are broken, it is expected that these two states can be adiabatically connected - they belong to the same phase.

{\bf Concluding remarks}:
In conclusion, RVL and RVB states, both of which are $Z_2$ quantum spin liquids, are two distinct phases protected by the given symmetry ($SO(3)$ spin-rotation or time-reversal). We have explicitly revealed the distinction between RVL and RVB states by studying the symmetry quantum numbers of their quasiparticles. This points to a systematic approach to (partially) classify symmetric topological phases in dimensions greater than one provided the generic interplay between symmetry quantum numbers and topological order (such as fusion rules of quasiparticles) is understood.  

{\bf Acknowledgment}:
We thank Sylvain Capponi, Zheng-Cheng Gu, Steve Kivelson, Dung-Hai Lee, Michael Levin, Masaki Oshikawa, and Xiao-Gang Wen for helpful discussions. We sincerely thank the hospitality of Institute of Physics in Beijing where this work was initiated, as well as KITP at UCSB where a part of the manuscript was written. This work is supported in part by DOE grant DE-AC02-05CH11231 at Berkeley (HY), the Harvard Society of Fellows (FL), and Alfred 
 P. Sloan Foundation (XLQ).


\begin{thebibliography}{21}
\bibitem{spinon} Note that spinon excitations have dynamics in our model.
\end{thebibliography}
\end{document}